\newcommand{\PRVL}[3]{#3 \ {\it Phys.\ Rev.\ Lett.}\ {\bf #1} \ #2}
\newcommand{\RMDP}[3]{#3 \ {\it Rev.\ Mod.\ Phys.}\ {\bf #1} \ #2}

\newcommand{\NAT}[3]{#3 \ {\it Nature}\ {\bf #1} \ #2}
\newcommand{\SC}[3]{#3 \ {\it Science}\ {\bf #1} \ #2}
\newcommand{\NATPHYS}[3]{#3 \ {\it Nature Phys.}\ {\bf #1} \ #2}

\newcommand{\PRA}[3]{#3 \ {\it Phys.\ Rev.\ A}\ {\bf #1} \ #2}

\newcommand{\PRE}[3]{#3 \ {\it Phys.\ Rev.\ E}\ {\bf #1} \ #2}

\newcommand{\PRD}[3]{#3 \ {\it Phys.\ Rev.\ D}\ {\bf #1} \ #2}
\newcommand{\JOPB}[3]{#3 \ {\it J.\ Phys.\ B:\ At.\ Mol.\ Opt.\ Phys.}\ {\bf #1} \ #2}
\newcommand{\JOPA}[3]{#3 \ {\it J.\ Phys.\ A:\ Math.\ Gen.}\ {\bf #1} \ #2}

\newcommand{\NJPH}[3]{#3 \ {\it New\ J.\ Phys.}\ {\bf #1} \ #2}

\newcommand{\PHYSREV}[3]{#3 \ {\it Phys.\ Rev.}\ {\bf #1} \ #2}

\newcommand\sech{\mathrm{sech}}
\newcommand{\SCR}{Schr\"odinger~}
\newcommand{\GP}{Gross-Pitaevskii~}
\newcommand{\diracslash}[1]{#1\llap{/\kern2pt}}

\newcommand{\be}{\begin{equation}}
\newcommand{\ee}{\end{equation}}
\newcommand{\bea}{\begin{eqnarray}}
\newcommand{\eea}{\end{eqnarray}}
\newcommand{\ba}[1]{\begin{array}{#1}}
\newcommand{\ea}{\end{array}}

\documentclass[12pt]{iopart}
\usepackage{graphicx}
\usepackage{verbatim}
\usepackage{lipsum}
\usepackage[pagebackref=false]{hyperref}
\begin{document}

\title{Bell Soliton in Ultra-cold Atomic Fermi Gas}

\author{Ayan Khan}\ead{ayankhan@fen.bilkent.edu.tr}
\address{Department of Physics, Bilkent University, 06800 Ankara, Turkey}
\author{Prasanta K. Panigrahi}\ead{pprasanta@iiserkol.ac.in}
%\authorrunning{A. Khan et. al.}
 \address{Indian Institute of Science Education and Research-Kolkata, Mohanpur, Nadia-741252, India}

\date{\today}

\begin{abstract}
%\abstract{
We demonstrate 
the existence of supersonic bell soliton 
in the Bardeen-Cooper-Schrieffer-Bose-Einstein condensate (BCS-BEC) crossover regime.
Starting from the extended Thomas-Fermi density functional theory of superfluid order parameter, a density
transformation is used to map the hydrodynamic mean field equation to a Lienard type equation. As a result, bell solitons
are obtained as exact solutions, which is further verified by the numerical solution of the
dynamical equation.
The stability of the soliton is established and its behavior in the entire crossover domain is 
obtained. It is found that, akin to the case of vortices, the bell solitons yield highest contrast in the BEC regime.
\end{abstract}

\maketitle
\section{Introduction} 
The experimental observation \cite{ketterle1} of superfluidity in the   
Bardeen-Cooper-Schrieffer-Bose-Einstein condensate (BCS-BEC) crossover regime
has given impetus to the study of ultra-cold trapped Fermi gases at unitarity \cite{stringari1,bloch}.
Though the beginning of the experimental study of superfluidity in 
cold atomic gases dates back to the observation of BEC, 
a decade and half ago \cite{cornell1,ketterle2}, the interaction 
controlled superfluidity, where atomic
Fermi gas changes constituents from Cooper pairs to composite bosons,
is fairly new \cite{regal}. As the atoms pass through a situation, where
the inter-atomic s-wave scattering length diverges (the unitarity), several novel phenomena
have been observed, some having strong resemblance with 
characteristics of high-$T_{c}$ superconductivity \cite{cho}.
The theory of crossover (the interaction driven study from BCS to BEC passing the unitarity) was first suggested long time back \cite{eagles} and 
has been studied systematically in the early eighties \cite{legget,smithrink}. 
The recent advancement in experiments \cite{jin1,jin2,chin,ketterle3,ketterle4} has
opened up avenues for deeper understanding of nonperturbative aspects of this strongly interacting
system, in a controlled manner.

Lower dimensional nonlinear systems are known to exhibit counter intuitive features, which are not possible in the 
three dimensional world. A classic example being existence of bright and dark 
solitons in quasi one dimensional BEC \cite{carr1,carr2,jackson,komineas,khyakovich,stecker,khawaja,burger,denschlag,cornish}, 
which derive their stability from 
the balancing of nonlinear effect with dispersion. The spreading effect of dispersion is exactly compensated by 
non-linearity, resulting in these stable structures \cite{shabat}.
Here, we show the existence of bell shaped solitons (bell-antibell pair) in the
ultra-cold atomic gases which are non-topological in nature. 
%%%%%%%%%%%%%%%%%%%%%%%%%%%%%%%%%%%%%%%%%%%%%%%%%%%%%%%%%%%%%
We note that a soliton solution can be classified in many ways, depending upon its shape (bell, kink, breather); nature of the non-linear systems (thus governing non-linear equations) which
lead KdV (pulse) soliton, sine-Gordon soliton, non-linear \SCR (NLS) soliton (envelop soliton); depending on its topology such as topological and non topological solitons etc.
Contrary to the envelop (bright) solitons, a Bell soliton is generally obtained from a KdV type of equation and the amplitude of the soliton is directly proportional to its velocity \cite{zhang}.
%%%%%%%%%%%%%%%%%%%%%%%%%%%%%%%%%%%%%%%%%%%%%%%%%%%%%%%%%%%%%
Similar hyperbolic secant pulse profiles have been observed in a number of natural systems, 
the most notable being the phenomenon of
self-induced transparency in atomic media \cite{hahn}. We establish the conditions under which such localized structures can manifest 
in the fermionic cold gases. This is explicated through exact solution in a controlled mean field approach, which enables
us to observe the pulse dynamics, starting from the fermionic side and ending up in the Bose-Einstein condensate.
 
In the context of the BCS-BEC crossover, mean-field microscopic description is well captured by the 
Bogoliubov-de Gennes (BdG) equation \cite{gennes}. The analysis of BdG equation is considerably 
more involved as compared to the \GP (GP) equation,
describing the mean field dynamics of BEC.
Antezza {\it et al.}, have numerically solved the BdG equation in a box and discussed the behavior of dark soliton, 
in a superfluid Fermi gas along the entire crossover domain, in a quasi-one dimensional (Q1D) geometry \cite{stringari2}. 
Similar approach was also adopted in the later studies as well \cite{strinati,scott}.
Manini {\it et al.}, have developed a simpler approach, using extended Thomas-Fermi density functional theory (ETFDFT), 
which incorporates the interaction 
through polytropic density of state \cite{salasnich1}, where
it was shown that for $^6Li$, it is possible to accurately find the collective properties in the 
crossover regime. 
More recently, Wen {\it et al.}, have presented the dark and bright soliton solutions
through the route of perturbative analysis of the above mean field equation \cite{wen1,wen2}. This leads to Korteweg-de Vries (KdV) equation 
whose soliton solutions are well known. 

Before going into the details of our analysis, here we like spend some time on the pros and cons of the ETFDFT model.
As mentioned before, a BdG analysis is one of the most robust way of treating the unitary Fermi gas as it takes into account the main contribution to the
kinetic energy, and treat it exactly in noninteracting systems even with a nonuniform density spatial variation. Where as this approach gives the exact kinetic energy only for a uniform
system and even when extended with the addition of gradient
and higher order derivatives of the density, the ETF functional is not able to reproduce shell effects in the density profile. But in this model one need not to worry about the
number of particles which is a big concern in BdG approach. More over through ETFDFT one can obtain the qualitative physical picture at zero temperature with relative numerical ease. 
Thus it is worth looking at the unitary Fermi system using this model.
 
Here, we present (i) an exact bell soliton solution from
the ETFDFT by showing a direct connection of ETFDFT with Lienard type of equations. Thereby we obtain the soliton solution which does not necessitate any perturbative approach \cite{wen1,wen2}.
These solitons have highest
contrast in the BEC regime, similar to the earlier result for vortices and dark solitons \cite{stringari2,randeria}.
(ii) We explore the collective excitations and obtain the sound velocity which shows an explicit dependence 
on the phenomenological $\lambda$ parameter (unless we absorb the quantum pressure term inside the chemical potential).
(iii) The stability of these solitons are discussed in terms of Noether charge. In the last few passages we present (iv) a scheme to treat ETFDFT analytically 
when the ultracold atomic Fermi gas is subjected to a harmonic trap and we observe the resulting oscillating dynamics of the soliton due to the parabolic confinement.

The article is organized as follows: In Sec.~\ref{poly}, we discuss the polytropic approximation 
for obtaining the relevant equation of state in the crossover regime.
A brief account is also presented on dimensional reduction of mean-field dynamics, to deal with quasi 1D scenario. 
In Sec.~\ref{loca}, we have presented the exact connection between the ETFDFT and 
Lienard type equation, from which we obtain bell soliton solutions.
We then explore its evolution from 
weak to strong coupling regimes. Further, we present the oscillating dynamics of the soliton and establish its supersonic nature by
analyzing the sound velocity in this medium. The stability is analyzed in the light of Noether charge and associated 
energy. Sec.~\ref{con} is devoted for the concluding remarks and
future directions of study.

\section{Polytropic Approximation and Quasi-One-Dimensional Dynamics}\label{poly}
\subsection*{A. Interaction in Unitary Fermi Gas}
In recent years, Manini {\it et al.}, have shown a simple way to investigate 
the unitary Fermi gas without losing its essential essence \cite{salasnich1}. In particular, for $^6Li$ they have built an efficient parametrization of the 
energy per particle, based on Monte Carlo data and asymptotic behavior of the interaction. Prior to that, it was 
shown that a power law dependence of equation of state can be effective to find reliable expressions for the 
collective frequencies 
%%%%%%%%%%%%%%%%%%%%%%%%%%%%%%%%%%
which is analytically treatable but preserves the vital physical picture \cite{stringari3}. Later it was demonstrated that,
at zero temperature under local density approximation the homogeneous chemical potential can be written in a series expansion of gas parameters with good effect \cite{ash}.
%%%%%%%%%%%%%%%%%%%%%%%%%%%%%%%%%%% 

Here, we make use of this ETFDFT
approach to investigate the existence of solitonic modes in superfluid 
Fermi gas.
We start from the ETFDFT equation, which controls the dynamics of the trapped superfluid Fermi gas
\cite{salasnich2}:
\begin{eqnarray}
 i\hbar\frac{\partial}{\partial t}\Psi & = & [-\frac{\hbar^{2}}{4m}\overrightarrow{\nabla}^{2}+2U(\mathbf{r})+2g(|\Psi|^{2})%\nonumber\\
+(1-4\lambda)\frac{\hbar^{2}}{4m}\frac{\overrightarrow{\nabla}^{2}|\Psi|}{|\Psi|}]\Psi,\label{eq:2}
\end{eqnarray}
where $\Psi$ is the Ginzburg-Landau order parameter at zero temperature and the
normalization condition $\int d\mathbf{r}|\Psi|^2=N$ ($N$ being the total atomic pair number of the superfluid Fermi gas) 
and $U(\mathbf{r})$ is the 
trapping potential.
The bulk chemical potential or equation of state is characterized by $g(n)$. We can introduce
the polytropic approximation for equation of state as $g(n)=\mu_{0}n^{\gamma}$. 
%%%%%%%%%%%%%%%%%%%%%
As discussed earlier, the polytropic approximation holds good at zero temperature under local density approximation which is essentially a perturbative approach capable of
providing deep insight into the physical nature of the quantum system \cite{ash}.
%%%%%%%%%%%%%%%%%%%%%
Under the polytropic approximation, $\mu_{0}$ is the reference
chemical potential, which we will explain after a short while, and $n$ is the atomic density. 
It is worth noting that, the interaction term is always positive in the whole BCS-BEC crossover.
We consider $\lambda$ as a phenomenological parameter, accounting for the increase of kinetic energy due to 
the spatial variation of the density. 
%%%%%%%%%%%%%%%%%%%%%%%%%%%%%%%%%%%%%%

The motivation to introduce the $\lambda$ comes from the density functional theory where a gradient correction was introduced to account for the non uniformity of the system \cite{salasnich4}.
An energy functional for fermions in terms of the density and its derivatives is usually
called extended Thomas-Fermi (ETF) functional.
Now if we define $n(\mathbf{r})$ as the local number density then a quantity like $-\lambda\hbar^2\nabla^2\sqrt{n}/(2m\sqrt{n})$ can be interpreted as a next to leading term correction factor, where $m$
being the atomic mass.
%%%%%%%%%%%%%%%%%%%%%%%%%%%%%%%%%%%%%%
There exits some ambiguity regarding the value of $\lambda$ as well. 
Initially the value was considered as $1$ in context of BCS-BEC crossover \cite{salasnich1} but later 
it was suggested that $\lambda=1/4$ is better choice based on its good agreement with the theoretical epsilon analysis around $d=4-\epsilon$ spatial dimensions in unitary regime \cite{salasnich4}.
For $\lambda=1/4$, Eq.(\ref{eq:2}) reduces to a form similar to the GP equation. 
%%%%%%%%%%%%%%%%%%%%%%%%%%%%%%%%%%%%%%%%%%%%%%%%%%%%%%%%%%%%%%%%%%%%%%%%%%%%%%%%%%%%%%%%%%
Since the formation of soliton is due to the delicate balance between dispersion and nonlinearity, 
and in quasi one dimensional (Q1D) systems, the quantum pressure has significant contribution to dispersion, therefore
one must pay careful attention to $\lambda$. Further, in Q1D the particle number of the condensate is small, 
which makes the contribution of quantum pressure more significant. 
%However, the situation can change if one 
%considers large number of atoms in the system, wherein the pressure effect is not significant. 
%Here, for generality, we carry out the calculation with arbitrary $\lambda$ and at the end we also present our result for $\lambda=1/4$,
%for explicating the effect of quantum pressure.
For large number of atoms, the radial contribution becomes significant. The ground state wave function in the radial direction
is Thomas-Fermi (TF) type, which results in discrete eigenmodes of linear excitations in the 
radial direction, and hence induces a dispersion effect on the wave propagation in axial direction.
The resulting dispersion can balance the nonlinear interaction and make the system more favorable to form 
solitons.

The essence of the interaction driven crossover is embedded in the polytropic coefficient $\gamma$.
Therefore, before going into the details of calculation, we briefly summarize the behavior of $\gamma$ and associated 
reference chemical potential $\mu_{0}$.
Here, the weak coupling to strong coupling evolution is captured through the following relation \cite{salasnich1},
\begin{eqnarray}
 \gamma&=&\frac{\frac{2}{3}\epsilon(y)-\frac{2y}{5}\frac{\partial\epsilon(y)}{\partial y}+
\frac{y^2}{15}\frac{\partial^2\epsilon(y)}{\partial y^2}}{\epsilon(y)-
\frac{y}{5}\frac{\partial\epsilon(y)}{\partial y}},\label{eq:1}
\end{eqnarray}
where $\epsilon(y)$ is an analytical function, whose parameters are fixed using the Monte 
Carlo data \cite{giorgini}, and $y=(k_{F}a)^{-1}$.
%%%%%%%%%%%%%%%%%%%%%%%%%%%%%%%%%%%%%%%%%%%%%%%%%
If one plots Eq.~(\ref{eq:1}) one will find that $\gamma\in[2/3,1]$ (roughly), while $y\in[-1,1]$, thus at $y=1$ Eq.~(\ref{eq:2}) effectively reduces to the \GP equation.
Where as, in the Fermi system the ground state energy is proportional to (density)$^{2/3}$ therefore, one can obtain the BCS like picture while $\gamma\simeq 2/3$ and thereby
it becomes possible to retain the essence of both the limits.
%%%%%%%%%%%%%%%%%%%%%%%%%%%%%%%%%%%%%%%%%%%%%%%%%
In a similar fashion, the reference chemical potential can be defined as, 
\begin{eqnarray}
 \mu_{0}&=&\epsilon_{F}\left[\epsilon(y)-\frac{y}{5}\frac{\partial\epsilon(y)}{\partial y}\right].\label{mu0}
\end{eqnarray}

The study of the superfluids at zero temperature can take various routes. One simple yet effective way is to study the 
hydrodynamics of the system. 
The ground state solution of superfluid hydrodynamic equations corresponds to 
$\partial n_{s}/\partial t=0$ and $\mathbf{v}_{s}=0$ ($n_{s}$, $\mathbf{v}_{s}$ being the superfluid density and velocity components)\cite{wen2}. As a result, it is possible to determine the ground state chemical potential
($\mu_{g}$) of the superfluid system: $\mu_{0}(n_{q}(\mathbf{r}))+U(\mathbf{r})=\mu_{g}$.
Here $n_{q}$ denotes equilibrium density, which can be related to the reference density through the polytropic approximation
as $n_{q}=n_{0}[\frac{\mu_{g}-U(\mathbf{r})}{2\mu_{0}}]^{1/\gamma}$, 
$n_{0}$ is particle number density.
One can obtain $\mu_{0}$ from Eq.(\ref{mu0}) and $n_{0}$ is taken as 
the density of the ideal Fermi gas at the trapping center of the system. 
The external potential is the usual harmonic trap, which in cylindrical coordinate takes the form, 
$U_{x,y,z}=\frac{1}{2}m(\omega_{\perp}r^2+\omega_{z}^2z^2)$, 
with $r=\sqrt{x^2+y^2}$.
The TF radius in radial and axial directions can be written as, $R_{xy}=2\mu_{g}/m\omega_{\perp}^2$ and
$R_{z}=2\mu_{g}/m\omega_{z}^2$ respectively. 
Applying the normalization
condition for equilibrium density ($\int n_{q}d\mathbf{r}=N$) the ground state chemical potential can be written as \cite{wen2},
\begin{eqnarray}
\mu_{g}&=&2\epsilon_{F}\left[\left\lbrace\epsilon(y)-\frac{y}{5}\frac{\partial\epsilon(y)}{\partial y}\right\rbrace^{1/\gamma}%\nonumber\\
\times\frac{\sqrt{\pi}(1+\gamma)\Gamma(\frac{1}{\gamma}+\frac{5}{2})}{8\gamma\Gamma(\frac{1}{\gamma}+2)}\right]^{2\gamma/(2+3\gamma)},\nonumber
\end{eqnarray}
where $n_{0}=(2m\epsilon_{F})^{3/2}/6\pi^2\hbar^3$ and $\epsilon_{F}=\hbar(6N\omega_{\perp}^2\omega_{z})$. In the 
preceding sections, it will be established that $\mu_{g}$ plays a crucial role in shaping the dynamics of soliton and also sound propagation
in this medium.
\subsection*{B. Dynamics in Quasi One Dimension}
To model a cigar shaped trap, it is essential to consider that the interaction energy in 
the transverse direction is much less than the kinetic energy \cite{salasnich3}. 
The fermions are assumed to be confined in a cylindrical harmonic
trap, with radial frequency much larger than the axial one, i.e., $\omega_{\perp}>>\omega_{z}$. The important assumption we need to make for employing the Gaussian variational ansatz in BEC in such situations is that, the interaction energy per particle must be weaker than the zero point energies associated with the harmonic confinement \cite{pethick}. Though this method has its own limitation, it is widely used in dilute Bose gas. Since the current ETFDFT model also lies on dilute limit where the inter-particle separation is much smaller than the $s$-wave scattering length therefore we are safe to use the similar Gaussian ansatz here as well.

To reduce Eq.(\ref{eq:2}) into Q1D form by using Gaussian ansatz, we consider
\begin{eqnarray}
 \Psi(\mathbf{r},t)&=&\frac{\sqrt{N}}{\sqrt{\pi a_{\perp}^2\sqrt{\lambda}}}\Phi\left(\frac{z}{a_{\perp}},\omega_{\perp}t\right)
\times%\nonumber\\
\exp{\left(-i\sqrt{\lambda}\,\omega_{\perp}t-\frac{x^2+y^2}{2a_{\perp}^2\sqrt{\lambda}}\right)},\label{eq:3}
\end{eqnarray}
where $a_{\perp}=\sqrt{\hbar/m\omega_{\perp}}$.
The reduced nonlinear \SCR equation (NLSE) then becomes,
\begin{eqnarray}
i\frac{\partial\Phi}{\partial t} & = & -\frac{1}{4}\frac{\partial^{2}\Phi}{\partial z^{2}}
+\omega_{0}z^{2}\Phi+2\mu_{0}(|\Phi|^{2})\Phi%\nonumber\\
+(1-4\lambda)\frac{1}{4}\frac{\frac{\partial^{2}|\Phi|}{\partial z^{2}}}{|\Phi|}\Phi,\label{eq:4} 
\end{eqnarray}
where $\omega_{0}=\omega_{z}/\omega_{\perp}$, and $\mu_{0}$ is appropriately
normalized by $\hbar\omega_{\perp}$.

We first analyze the results without the trap, and later we will incorporate 
the trap to obtain analytic solutions. 
The solution of the above NLSE, without the pressure and trap have been studied in different perspective \cite{khare1,lin}.
Using a traveling wave ansatz of the type $\Phi(z,t)=\rho(z,t)\exp{i(\chi(z,t)+\mu_{g}t)}$, we study the system in  
the comoving frame, $\zeta=z-Vt$, where $V$ is the soliton velocity. 
After applying these transformation in Eq.(\ref{eq:4}), the imaginary
equation (imaginary part of Eq.(\ref{eq:4})) arising from the current conservation gives,
\begin{eqnarray}
 \chi_{\zeta}=\frac{4C_{0}}{\rho^2}+2V,\label{eq:42}
\end{eqnarray}
where $C_{0}$ is an integration constant. We consider the phase and amplitude as uncorrelated and hence
assume $C_{0}=0$. The real equation (real part of Eq.(\ref{eq:4})) then reads,
\begin{eqnarray}
\lambda\rho_{\zeta\zeta}+(V^2-\mu_{g})\rho-2\mu_{0}\rho^{2\gamma+1}=0,\label{eq:43}
\end{eqnarray}
 where we have used the results from Eq.(\ref{eq:42}). The subscript $\zeta$ means derivative with respect to $\zeta$.
It is worth noting that for $\gamma=1$,
one can have full set of Jacobi elliptic function as solutions. 
For Fermi gas, we do not have such freedom and we need to solve the equation, considering $\gamma$ as
unknown. Later on we will use selected values of $\gamma$ obtained from Eq. (\ref{eq:1}).
%%%%%%%%%%%%%%%%%%%%%%%%%%%%%%%%%%%%%%%%%%%%%%%%%%%%%%%%%%%%%%%%%%%%%%%%%%%%%%%%%%%%%%%%%%%%%%%%%%%%%%%%%%%%%%%
%%%%%%%%%%%%%%%%%%%%%%%%%%%%%%%%%%%%%%%%%%%%%%%%%%%%%%%%%%%%%%%%%%%%%%%%%%%%%%%%%%%%%%%%%%%%%%
\section{Localized Structure in the Crossover Regime}\label{loca}
\subsection*{A. Bell Soliton}
Till date only a few studies have addressed the complexity involving the polytropic assumption.
Very recently, Wen {\it et al.} \cite{wen1,wen2}, made such efforts and studied both dark and bright solitons using a
perturbative expansion scheme on the mean field equation.
Interestingly, there exists a transformation which will enable us to reduce 
this equation to a well known form, which has exact solutions. 
Without loss of generality one can take $\rho=\sigma^{1/\gamma}$, 
which connects Eq. (\ref{eq:43}) to a Lienard type equation:
\begin{eqnarray}
 \frac{1}{\gamma}\sigma\sigma_{\zeta\zeta}-\frac{\gamma-1}{\gamma^2}\sigma_{\zeta}^2+P_{1}\sigma^2-P_{2}\sigma^4=0,\label{eq:11}
\end{eqnarray}
where $P_{1}=(V^2-\mu_{g})/\lambda$ and $P_{2}=2\mu_{0}/\lambda$. We like to mention here that one can obtain a real constant solution from Eq.(\ref{eq:11}) iff $V^2-\mu_{g}>0$,
as $\sigma=\sqrt{P_{1}/P_{2}}$. But soon we will see that the prevailing condition is inconsistent while we try to obtain a localized solution.
Thus we refrain ourselves from the constant solution.
To obtain a localized solution we apply
fractional transformation \cite{lin} of the following form,
\begin{eqnarray}
 \sigma&=&\frac{A\,\sech^2[B \zeta/2]}{1+C\,\sech^2[B \zeta/2]}.\label{eq:12}
\end{eqnarray}
As short recap, we like to remind the readers that fractional transformation is an useful tool to obtain solutions from a non-linear equation. In this approach, one can consider 
an ansatz solution in the following way,
\begin{eqnarray}
\sigma(\zeta)=\frac{A+Bf^{\delta}(\zeta)}{C+Df^{\delta}(\zeta)}\nonumber,
\end{eqnarray}
where $AD-BC\neq0$, $\delta$ can be any power of the function $f$. $A$, $B$, $C$, $D$ are constant which can be evaluated through consistency conditions after applying this ansatz in the
main equation and collecting the coefficients according to the power of $f$. 
A comprehensive description of this process can be found in Ref. \cite{raju}. 

We follow the same prescription, thereby application of Eq.(\ref{eq:12}) in Eq.(\ref{eq:11})
leads to the following set of consistency conditions:
\begin{eqnarray}
-16\gamma P_{1}+16B^2\gamma-32B^2 & = & 0,\nonumber\\
-2C\gamma P_{1}-16B^2\gamma+B^2C+40B^2&=&0,\nonumber\\
\textrm{and}\,\,A^2\gamma P_{2}-C^2\gamma P_{1}-8B^2C&=&0.\nonumber
\end{eqnarray}
The obtained solutions are,
\begin{eqnarray}
A=\pm\frac{1}{2}\sqrt{\frac{\gamma P_{1}-3P_{1}}{\gamma P_{2}-2P_{2}}},B&=&\pm\sqrt{\frac{\gamma P_{1}}{\gamma-2}}
\,\,\textrm{and}\,\,C=-1/2.\nonumber
\end{eqnarray}
%%%%%%%%%%%%%%%%%%%%%%%%%%%%%%%%%%%%%%%%%%%%%%%%%%%%%%%%%%%%%%%%
\begin{comment}
 The final solution irrespective of the interaction can be written as,
\begin{eqnarray}
 \rho(T)&=&\left\lbrace2\sqrt{\frac{(V^2-\mu_{g})(\gamma-3)}{\mu_{0}\alpha(\gamma-2)}}\sech\left[\sqrt{\frac{\gamma (V^2-\mu_{g})}{\gamma-2}}\zeta\right]\right\rbrace^{1/\gamma}.\label{eq:13}
\end{eqnarray}
%A hyperbolic identity ($\cosh{z/2}=\left[\frac{\cosh{z}+1}{2}\right]$) \cite{stegan} is applied to 
%write our solution in the above form.
It is possible to extract a soliton solution for ultracold Fermi gases in presence of trap
which is a bell soliton type. It should be reiterated that using Eq.(\ref{eq:6}), it is possible to deal with 
different type of trapping potentials as per requirement. 
\end{comment}
%%%%%%%%%%%%%%%%%%%%%%%%%%%%%%%%%%%%%%%%%%%%%%%%%%%%%%%%%%%%%%%%%%%%
\begin{figure*}
\begin{center}
%\begin{minipage}[c]{7.cm}
\includegraphics[width=5.cm]{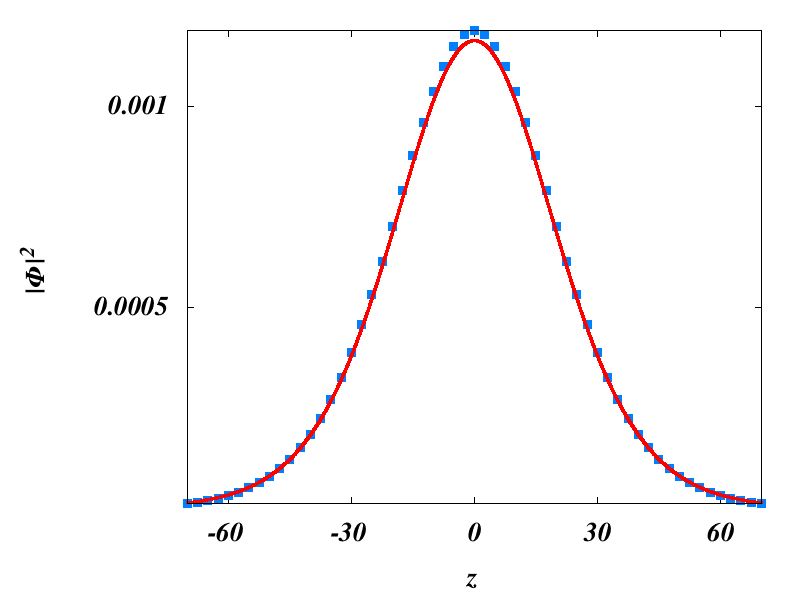}
\includegraphics[width=5.cm]{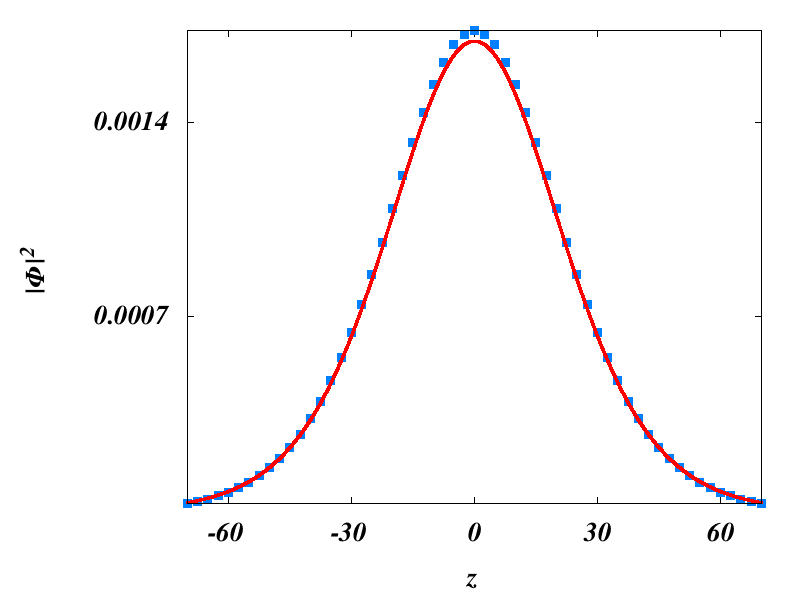}
\includegraphics[width=5.cm]{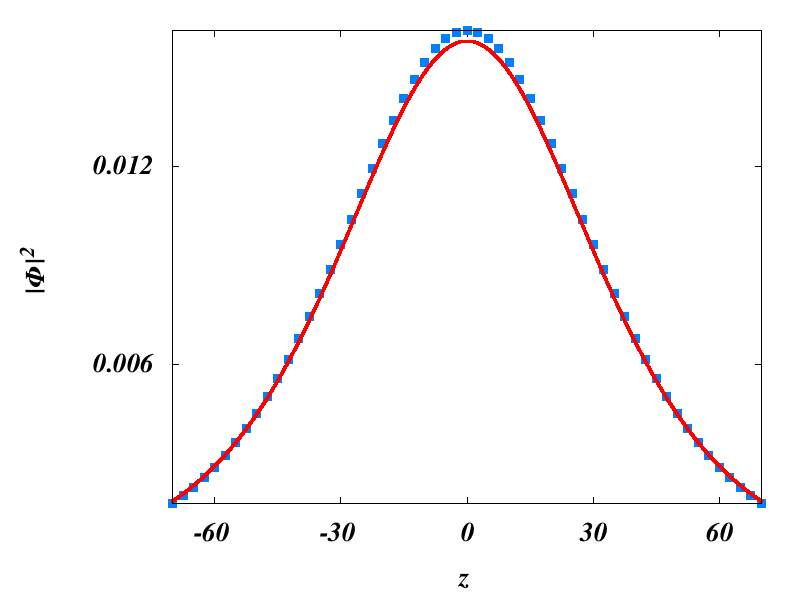}
\caption{(Color online)The bell soliton profiles across the BCS-BEC crossover regime is depicted. The interaction is tuned from weak to strong
($1/k_{F}a=-1$, $0$, $1$) from top to bottom. The red line is our analytic result and the blue squares denote corresponding numerical 
solutions after evolution of the Q1D ETFDFT equation.}\label{rho}
%\end{minipage}
\end{center}
\end{figure*}

Since our motivation is
to find localized solutions, it is important to impose $\mu_{g}-V^2>0$ (contrary to the constant solution), otherwise 
the solution will be of trigonometric form
(since $\gamma-2<0$ always for this kind of system) having singularities in time evolution. 
One should note that, the above assumption actually imposes a ceiling on the soliton velocity contrary to the usual bright soliton solution whose velocity is not bounded.
Incorporating this assumption, the solution can be simplified to the compact form (after normalization),
\begin{eqnarray}%\flushleft
\rho(\zeta)&=&\left\lbrace\frac{\eta e^{i\pi/2}}{2\sqrt{B(-1,\frac{1}{\gamma},\frac{\gamma-2}{\gamma})}} 
\sqrt[4]{\frac{\gamma(V^2-\mu_{g})}{\lambda(\gamma-2)}}\right\rbrace^{1/\gamma}%\nonumber\\
\times\left\lbrace\sech\left[\sqrt{\frac{\gamma (V^2-\mu_{g})}{\lambda(\gamma-2)}}\zeta\right]\right\rbrace^{1/\gamma},\label{eq:14}
\end{eqnarray}
where $\eta=\pm1$ is the polarity of the soliton (bell/anti-bell) and $B(-1,\frac{1}{\gamma},\frac{\gamma-2}{\gamma})$ is the incomplete beta function.
It is worth pointing out that, this is a distinct and new solution (bell type), as compared to the solution presented in Ref.~\cite{wen2}.
Fig.(\ref{rho}) demonstrates the bell soliton at time $t=0$, for different couplings, with 
$V^2=0.9\mu_{g}$. The solid red line is our analytic result (Eq. (\ref{eq:14})), the corresponding numerical solution
is depicted through blue squares. The numerical result presented here are obtained via solving ETFDFT equation
using the split-step Crank-Nicolson method by discretizing in space and time \cite{adhikari}. The discretized equation is then solved by
propagation in imaginary time over small time steps ($0.00002$, total time step is taken as $8000$). 
Both the results agree appreciably well with each other. 

One must notice that in Eq.(\ref{eq:14}), if $\gamma=2$,
the solution will be unstable as referred in Ref.~\cite{khare1} but in our case on physical grounds $\gamma<2$
(though mathematically $\gamma$ can have any value but in this model $\gamma$ values are fixed by Eq.(\ref{eq:1}), thus its value can never exceed $1$).
It is important to observe the change in visibility of the bell solitons across the crossover.
The visibility/amplitude of the bell soliton is different in different superfluid regimes. For dark soliton, 
it has already been shown that visibility is maximum in the BEC side \cite{stringari2}. 
In Fig.(\ref{rho}), we notice that the visibility does increase as we move from weak coupling to 
strong coupling regime. Further, one can note that the change in visibility is more rapid in BEC side
and this increase is monotonic.
A qualitative comparison with the visibility of vortex core density in the BCS-BEC 
crossover \cite{randeria} reveals striking resemblance with our study.

\subsection*{B. Sound Velocity and Stability Criterion}
Before presenting the sound velocity result in our system, we intend to address one intriguing issue regarding the sound velocity.
It is well know that one can calculate the sound velocity through the hydrodynamics as well as from the ETFDFT equation (Eq.\ref{eq:2})\cite{salasnich2,pethick}.
It is worth noting that the effect of quantum pressure does not enter in the sound velocity expression if we use the hydrodynamic approach ($c_{s}=\sqrt{\gamma\mu_{g}}$). This is even true
for ETFDFT approach if we absorb the quantum pressure term in the chemical potential. But otherwise the quantum pressure term will leave an effect on the sound velocity.

We now analyze the soliton propagation velocity, for the bell solitons and relate it to the sound velocity.
A general way to obtain the sound velocity is by applying a small perturbation about a constant background.
A systematic calculation reveals,
$c_{s}=\sqrt{2\lambda\mu_{0}\gamma}\rho_{0}^{\gamma}=\sqrt{2\lambda\gamma\mu_{g}}$ 
(since $\rho_{0}^{2\gamma}=\mu_{g}/\mu_{0}$) considering the homogeneous unperturbed condensate density varies slowly in space.
One should note here that the contribution arising from the quantum pressure 
($\lambda$) also makes a presence in determining the sound velocity.
Qualitatively our results are in good agreement with Ref.~\cite{wen2} but small
quantitative difference might be due the fact that, our analysis
is very different in nature from the previous analysis where collective excitations were calculated
from the density fluctuation in three dimension.
As an example, in the deep BEC limit with $\gamma=1$ and $\lambda=1/4$,
$c_{s}^{BEC}=\sqrt{\mu_{g}/2}$ as obtained in Ref.~\cite{wen2}.
We have seen that condition for existence of bell soliton is $V^2<\mu_{g}$. From the 
above discussions on sound, one can write $V^2<c_{s}^2/(2\lambda\gamma)$, which allows a wide range of 
soliton velocity. However, this model accepts a certain range of values for $\gamma$. More precisely, 
$\gamma\in[2/3,1]$. Hence it turns out that $V\in[1.73 c_{s},1.41 c_{s}]$, which points to 
the supersonic nature of the soliton velocity. 
\begin{table}[ht]
\caption{Comparison of different parameters from weak to strong coupling limit for $V^2=0.9\mu_{g}$.}\label{comparison}
\centering 
\begin{tabular}{c c c c c c c}
\hline\hline 
$\frac{1}{k_{F}a_{s}}$ & $\gamma$ & $\mu_{0}$ & $\mu_{g}$ & $c_{s}|_{\lambda=0.25}$ & $c_{s}|_{\lambda=0.13}$\\ % inserts table
%heading
\hline % inserts single horizontal line
-1 & 0.624 & 0.758 & 1.92 & 0.773 & 0.558\\ % inserting body of the table
0 & 0.666 & 0.420 & 1.65 & 0.741 & 0.534 \\
1 & 1.053 & 0.083 & 0.59 & 0.557 & 0.401 \\
\hline %inserts single line
\end{tabular}
\label{table:nonlin} % is used to refer this table in the text
\end{table}

To understand the origin of bell soliton generation, we must look back at the role of the
ground state chemical potential. 
Generaly dispersion relation can be written as, $\omega\propto k$ where $\omega$ and $k$ is the frequency and wave number (momentum)
of the system respectively. This is usually referred as material dispersion. But there exists another dispersion mechanism which is known as waveguide dispersion.
This dispersion actually contains the dimensional effect i.e $\sqrt{\omega^2-\omega_{d}^2}\propto k$ where $\omega_{d}$ depends on the dimension
of the system. In our system, though the trap is effectively cigar shaped, we have taken into account the 
radial contribution in the density through $\mu_{g}$. Thus the TF distribution in the ground state, incorporated through 
$\mu_{g}$, supports the ``waveguide dispersion'', as opposed to the normal 
``material dispersion'', and we consider this different dispersion mechanism supports the generation of bell solitons \cite{wen2}.

For completeness, we list the values of several relevant parameters explicitly for different couplings in
Table \ref{comparison}. 
All the parameters are already normalized
by the Fermi energy or Fermi velocity according to the relevance.
The gradual decrease of reference chemical 
potential, as well as ground state chemical potential from BCS to BEC regime is understandable as it is known that
the BCS state is highly dense/closely packed, whereas the composite bosons are sparsely distributed. The gradual decrease
of sound velocity, as one moves towards strong coupling regime also agrees with the experimental data \cite{joseph}. The quantitative analysis
of our result reveals that, the most suitable value for the phenomenological parameter turns out to be $\lambda=0.13$ (which has been proposed in Ref. \cite{salasnich2}).
We also note that the calculated sound velocity using $\lambda=0.13$, augurs well with the sound velocity obtained through mean-field BCS-BEC crossover formalism \cite{stringari5}.
 \begin{table*}[t]
 \caption{Comparison of conserved charge and associated energy of the bell solitons in the BCS-BEC crossover regime
for different soliton velocity ($\lambda=1/4$).}\label{QE}
\centering 
\begin{tabular}{c|c c|c c|c c|c c|c c}
\hline
\hline
$\frac{1}{k_{F}a_{s}}$&\multicolumn{2}{c|}{$V^2=0.75\mu_{g}$}&\multicolumn{2}{c|}{$V^2=0.8\mu_{g}$}&\multicolumn{2}{c|}{$V^2=0.85\mu_{g}$} 
&\multicolumn{2}{c|}{$V^2=0.9\mu_{g}$}&\multicolumn{2}{c}{$V^2=0.95\mu_{g}$}\\
%\hline
\cline{2-11}
& $\mathcal{Q}\sqrt{\mu_{g}}$ & $\mathcal{E}(\mathcal{Q})$ & $\mathcal{Q}\sqrt{\mu_{g}}$ & $\mathcal{E}(\mathcal{Q})$ & $\mathcal{Q}\sqrt{\mu_{g}}$ 
& $\mathcal{E}(\mathcal{Q})$ & $\mathcal{Q}\sqrt{\mu_{g}}$ & $\mathcal{E}(\mathcal{Q})$ & $\mathcal{Q}\sqrt{\mu_{g}}$ & $\mathcal{E}(\mathcal{Q})$\\
\hline
-1 & 23.893 & 29.8898 & 18.682 & 20.561 & 13.604 & 12.762 & 8.7001 & 6.577 & 4.051 & 2.164\\
0 & 40.154 & 43.6536 & 32.112 & 30.137 & 24.074 & 18.765 & 16.039 & 9.691 & 8.011 & 3.186\\
1 & 25.002 & 35.8255 & 22.615 & 25.688 & 19.870 & 16.730 & 16.559 & 9.142 & 12.124 & 3.254\\
\hline\hline
\end{tabular}
\label{table:nonlin} 
\end{table*}

To investigate the stability of these bell solitons, which are non-topological in nature, one should calculate 
the Noether charge defined as $\mathcal{Q}=\int|\rho|^2 dz$,
and the corresponding energy,
\begin{eqnarray*}
 \mathcal{E}(\mathcal{Q})&=&\int\left[\lambda\Big|\frac{\partial\rho}{\partial z}\Big|^2+
\mu_{0}(|\rho|^2)^{2\gamma}\right]dz.\label{eq:18}
\end{eqnarray*}
The stability of these
solitons demands that 
$\mathcal{E}(\mathcal{Q})<\sqrt{\mu_{g}}\mathcal{Q}$ \cite{lee,khare}.
After evaluating the charge and energy integrals numerically, we list our result in Table \ref{QE}.
Using these results and the criterion for stability we find that solitons are more stable in the high velocity regime (in the vicinity
of $V\simeq\sqrt{\mu_{g}}$). At relatively lower velocity ($V\simeq\sqrt{0.75\mu_{g}}$) the whole crossover is unstable
but gradual increase in the velocity stabilizes the system for different interaction regime. At $V=\sqrt{0.8\mu_{g}}$ interestingly one 
can see that, the crossover regime is stable where as, the two extreme regimes are unstable. Afterwards, 
($V\simeq\sqrt{0.85\mu_{g}}$), the entire crossover satisfy 
the stability criterion. Hence, the onset of stability happens faster in 
the unitary regime, which is sensible as different experiments already pointed out that crossover is more stable compared to BCS and 
BEC sides \cite{ketterle1}.
\subsection*{C. Coherent Controll of Bell Solitons}
We will now extend our analysis in presence of harmonic confinement. For that purpose we start from Eq. (\ref{eq:4}) with interaction coefficient 
being time dependent i.e $\mu_{0}\equiv\mu(t)$ as suggested in Ref.\cite{atre}.
Since we are interested in finding soliton solutions, a self similar ansatz of the following 
form can be employed \cite{atre,rajendran},
\begin{eqnarray}
 \Phi(z,t)&=&\sqrt{\alpha(t)}\rho[\alpha(t)(z-l(t))]\times%\nonumber\\
\exp{[i(\theta(z,t)+\chi\{\alpha(t)(z-l(t))\})]},\label{eq:5}
\end{eqnarray}
where the phase has the exact form, $\theta(z,t)=\mu_{g}\int_{0}^{t}\alpha(t')^2dt'-c(t)z^2$, and $\chi$ being the density dependent
complex phase of the solution profile. 
Application of the ansatz in the NLSE and separation 
of real and imaginary parts result in two equations, related to the continuity (imaginary part) 
and pressure (real part) equations.  

The following consistency conditions emerge from Eq. (\ref{eq:4}),
\begin{eqnarray}
 \frac{d\alpha(t)}{dt}&=&\alpha(t)c(t), \qquad c(t)l(t)+\frac{dl(t)}{dt}=V,\label{eq:8}
\end{eqnarray}
The second equation in Eq. (\ref{eq:8}) represents conservation of the velocity components, 
where $V$ being a conserved quantity having the dimension of velocity.

The equation of continuity leads to the following phase relation,
\begin{eqnarray}
 \chi_{T}&=&\frac{4C_{0}}{\rho^2}+2V,\label{eq:9}
\end{eqnarray}
where $T=\alpha(t)(z-l(t))$.
As before, to avoid the amplitude dependence on the phase, we consider the integration constant $C_{0}=0$.
\begin{figure*}
\begin{center}
\includegraphics[width=5.cm]{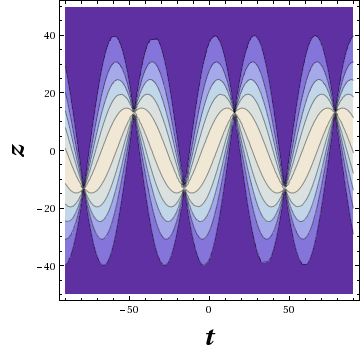}
\includegraphics[width=5.cm]{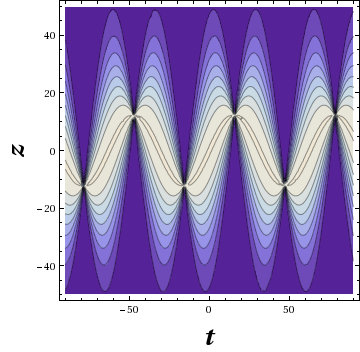}
\includegraphics[width=5.cm]{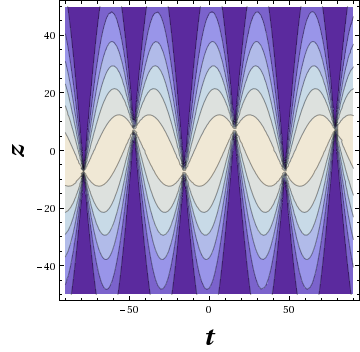}
\caption{(Color online) Time evolution of of the bell soliton shows oscillation inside the trap. The interaction
is tuned from weak to strong ($1/k_{F}a=-1$, $0$, $1$) depicted from left to right in the figure.}\label{rho0}
\end{center}
\end{figure*}

It is interesting to see that $c(t)$ satisfies a Riccati type equation,
$ \frac{dc}{dt}-c^2=\omega_{0}^2$,
which can be mapped to \SCR equation, through a transformation: $c(t)=-\frac{d\ln{\beta(t)}}{dt}$.  
This relation enables us to find the soliton profile for different type 
temporal variations of the trap frequency $\omega_{0}$ \cite{ranjani}.
At present, we will only demonstrate the effect for usual harmonic potential for which $\omega_{0}$
is constant.
This leads to the solution $c(t)=\omega_{0}\tan{[\omega_{0}t]}$ and $\alpha(t)=\alpha_{0}\sec[\omega_{0}t]$.
At this point, we can shed further light on the role of $V$. 
The presence of the trap necessitates chirping of the phase, which enforces another velocity component.
In the absence of the trap, it is appropriate
to consider $c(t)=0$, and then we can easily get back the well known velocity expression as, $l(t)=V t$.

The amplitude equation in the comoving frame leads to,
\begin{eqnarray}
 \lambda\rho_{TT}+(V^2-\mu_{g})\rho-2\tilde{\mu}(\rho^2)\rho=0.\label{eq:10}
\end{eqnarray}
The third term, involving the nonlinear interaction can now be written more precisely using polytropic approximation:
$2\tilde{\mu}(\rho^2)\rho=\mu_{0}\alpha\rho^{2\gamma+1}$, where $\tilde{\mu}=\frac{\mu(t)}{\alpha^2}$ and 
$\mu(t)=\mu_{0}\alpha^{3-\gamma}$. Thus the temporal variation of the interaction is coupled with width of the
solution, as was observed in in Ref.~\cite{atre}. It must be noted that Eq.(\ref{eq:10}) is same as Eq.(\ref{eq:43}).
Hence one can find the same solution with distributed coefficients (in the homogeneous case, $\alpha(0)=1$).

Fig.(\ref{rho0}) depicts the soliton oscillation at the entire crossover regime. One can find
similar type of solution in Ref. \cite{rajendran,ranjani1,ramesh} only for the BEC regime. Oscillation of soliton inside a trap in BEC 
is already 
a well documented fact \cite{atre,rajendran,weller,becker,busch,parker}. 
It has been concluded that the oscillation frequency of the soliton is 
$\frac{\omega_{z}}{\sqrt{2}}$. Recent study of dark soliton in the Fermi gas has predicted that
the oscillation in the unitary regime will be $\frac{\omega_{z}}{\sqrt{3}}$ \cite{stringari4,brand}.
In our analysis we consider $\omega_{0}=\omega_{z}/\omega_{\perp}=0.1$, 
which is reasonable if we keep in mind the actual experimental situations.
It should be mentioned that the oscillation frequency does not show any dependence on the inter atomic 
interaction, which might be due to the limitation of the hydrodynamic approach. 
\section{Conclusion}\label{con}
In precise, here we have demonstrated localized solitons in the Fermi gas 
of $^6Li$ by exploiting an analytical connection between ETFDFT and Lienard equation.
These solutions in quasi 1-D system have also been checked 
numerically, using split-step Crank-Nicolson method. 
The solitary waves draw support from the radial contribution
incorporated through $\mu_{g}$. 
The analysis of elementary excitations reveals  
the explicit dependence of sound velocity on the phenomenological parameter $\lambda$, unless we absorb the quantum pressure inside the chemical potential.
The propagation velocity of the 
bell soliton is supersonic and there exists a fixed upper bound 
of the velocity beyond which the system will be 
unstable. The upper bound is controlled by the ground state chemical potential in the equilibrium condition. 
We realize that the stability of solitons across the crossover is bounded in a narrow region of velocity,
approximately, $0.8\sqrt{\mu_{g}}<V<\sqrt{\mu_{g}}$. The lower bound is interaction dependent. It reveals that 
the stability is attained at unitarity at lower velocity than the other two regions. This matches well
with the recent finding that, the superfluid Fermi gas is more stable at unitarity \cite{ketterle1}. In the last part, we have shown 
the mathematical technique to tackle inhomogeneous ETFDFT analytically. 

In this article we have described how to obtain certain kind of localized solution in ultra cold Fermi gas using ETFDFT technique. 
Though this model has its own limitations (as mentioned before) but it allows you to make 
good qualitative idea about the strongly correlated Fermi system.
As a matter of fact, always one trades between simplicity and accuracy. Thus
we consider evolution of bell soliton through real space BdG analysis
using the polytropic density of state as an exciting next step in the near future.

As a concluding remark, we hope that these solitons can be experimentally verified in near future.
In experiments, solitons can be created in a similar way as
vortices are sometimes made \cite{ketterle1}. A bell soliton can be created on the
BEC side of the crossover and then the Feshbach resonance can be tuned towards unitarity or the
BCS side. Although vortices are made by imposing rotation on a BEC, bell solitons are made
via density engineering i.e a sudden switch on of a focused and blue-shifted laser beam should initiate a density flush thus producing bell soliton.
Current rapid growth as well as interest in ultracold atom research (specially in Fermi systems) encourage us to believe that
these solitons will be visible in the experiments soon.

\section*{Acknowledgement}
%\begin{acknowledgements}
AK would like to thank warm hospitality at IISER-Kolkata and the financial support from TUBITAK-
BIDEP and TUBITAK (112T176). AK also acknowledges interesting discussion with S. W. Kim,
S. Adhikari, F. Dalfovo and helpful communications with L. Salasnich, J. Brand and G. Huang.
%\end{acknowledgements}
%\begin{references}

%\end{references}

\end{document}